# Non-Orthogonal Multiple Access with Wireless Caching for 5G-Enabled Vehicular Networks

Sanjeev Gurugopinath, Sami Muhaidat, Yousof Al-Hammadi, Paschalis C. Sofotasios, and Octavia A. Dobre

*Abstract*—The proliferation of connected vehicles along with the high demand for rich multimedia services constitute key challenges for the emerging 5G-enabled vehicular networks. These challenges include, but are not limited to, high spectral efficiency and low latency requirements. Recently, the integration of cache-enabled networks with non-orthogonal multiple access (NOMA) has been shown to reduce the content delivery time and traffic congestion in wireless networks. Accordingly, in this article, we envisage cache-aided NOMA as a technology facilitator for 5G-enabled vehicular networks. In particular, we present a cache-aided NOMA architecture, which can address some of the aforementioned challenges in these networks. We demonstrate that the spectral efficiency gain of the proposed architecture, which depends largely on the cached contents, significantly outperforms that of conventional vehicular networks. Finally, we provide deep insights into the challenges, opportunities, and future research trends that will enable the practical realization of cache-aided NOMA in 5G-enabled vehicular networks.

## INTRODUCTION

The rapid development of wireless technologies has catalyzed the deployment of new paradigms such as connected vehicles (also dubbed as vehicular communication networks), which are considered key enablers for intelligent transportation systems (ITS). The design of these technologies must be inherently tailored to the stringent requirements of ITS services and their technological trends. In this context, the re-design of vehicular networking has gained enormous attention recently, bringing a new level of connectivity to vehicles and commercial potentials for their practical realization [1].

Vehicular networks comprise communication terminals that relay information and also exchange data with other entities. Potential applications of vehicular communications are diverse and pervasive. For example, transportation can be improved through fast dissemination of road and traffic information, as well as by enabling coordination of vehicles at critical points such as highway entries and intersections. In addition, several new applications can be facilitated, e.g., cooperative high-speed internet access from within the vehicular network, cooperative downloading, network gaming among passengers of adjacent vehicles, video-enabled meetings among co-workers travelling in different vehicles, and previously unimagined products which tend to be spawned by new communications services.

In order to realize these applications, a new communication paradigm, coined as vehicle-to-everything (V2X), has been recently proposed as a platform to enable communication between a vehicle and an entity. V2X incorporates several types of communications, such as V2I (vehicle-to-infrastructure), V2P (vehicle-to-pedestrian), V2N (vehicle-to-network), V2U (vehicle-to-user), and V2V (vehicle-to-vehicle) [1].

V2X communications can be realized based on two main technologies, namely WLAN-based and cellular-based networks. As a dedicated short-range communications (DSRC) technology, wireless access for vehicular environments, also known as 802.11p, provides high-speed V2V and V2I data transmission and enables data rates in the range of 6 - 27 Mbps over short distances [1]. On the other hand, the 3rd generation partnership project (3GPP) has recently published the cellular (C)-V2X specifications based on LTE as the underlying technology. Table I summarizes the characteristics and general properties of DSRC and C-V2X.[2]

In the recently proposed internet-of-vehicles (IoV) [2], the vehicles are envisioned to share and access enormous amount of data from data centers, causing increased latency and intermittent connectivity. Furthermore, due to the high mobility scenarios, network topologies can change quickly and handover becomes more frequent, presenting several additional challenges for the current vehicular networks, particularly in terms of reliability. Consequently, several 5G enabling technologies, such as multiple-input multiple-output (MIMO), heterogeneous network (HN)-based communications, millimetre-wave (mmWave) communications and ultra-wideband (UWB) communications are being considered in 5G-enabled vehicular networks.

TABLE I. COMMONALITIES AND DIFFERENCES BETWEEN DSRC AND C-V2X (SOURCE: WWW.AUTO-TALKS.COM).

| Feature | DSRC 802.11p | C-V2X Release 14/15 |
|---|---|---|
| Goal | Direct real-time communication between V2U and V2I | |
| Deployment | Since 2017 | Late 2020 |
| Cellular Connectivity | Hybrid model | Hybrid model |
| Communications Technology | OFDM with CSMA | SC-FDM with SP sensing |
| Security | Public key cryptography | |
| Infrastructure Investment | Dedicated camera and traffic light-based infrastructure | |
| Roadmap | 802.11bd to 802.11p | C-V2X Release 16 |

However, a key common challenge among these technologies is the significant amount of backhaul overhead [3]. In this respect, it has been recently shown that caching techniques, in which popular files are stored *a priori* at vehicles, are able to reduce the backhaul traffic by offloading traffic from the network. The basic idea of caching is to store popular contents in different geographical locations during off-peak time. This local storage arrangement enables a duplicate transmission of popular files, thereby increasing the

---

S. Gurugopinath, S. Muhaidat, Y. Al-Hammadi, and P. C. Sofotasios are with Khalifa University, UAE; S. Gurugopinath is also with PES University, India; O. A. Dobre is with Memorial University, Canada.

[2]The acronyms used are – orthogonal frequency division multiple access (OFDM), carrier-sense multiple access (CSMA), single-carrier frequency division multiplexing (SC-FDM), and semi-persistent (SP).

spectral efficiency and reducing latency. Additionally, it has been demonstrated in the literature that installing memory units at the user-end is indeed a cost-effective solution compared to increasing the backhaul overhead [3]. Consequently, several investigations on the integration of caching techniques with recent technologies, such as HNs, device-to-device communications, as well as cloud- and fog-radio access networks, have been considered in the recent literature.

Non-orthogonal multiple access (NOMA) is another key technology, which has been recently proposed as a promising multiple access solution to address some of the challenges in 5G networks [4]. Owing to its notable advantages, such as higher spectral efficiency and increased number of users, NOMA is envisioned as a potential solution to the problem of massive connectivity in 5G networks and beyond. In cellular networks, the advantages of NOMA in comparison with the conventional orthogonal multiple access (OMA) have been extensively studied, in which significant performance gains in terms of spectral and energy efficiencies have been reported [4], [5]. A NOMA system can be built either in (a) power-domain or (b) code-domain. In the power-domain NOMA (PD-NOMA), users are assigned different power levels over the same resources, while in the code-domain NOMA, multiplexing can be carried out using spreading sequences, similar to code division multiple access technology [6]. In this article, the term NOMA is specific to PD-NOMA. However, the techniques described here can also be extended to other classes of NOMA.

The key idea in a downlink-NOMA system is that vehicles rely on successive interference cancellation (SIC) for data detection. In particular, the vehicle that has been assigned the highest power, decodes its own signal by treating the interference from other signals as noise. Other vehicles progressively cancel out the decoded signal from those with higher power, and then decode their own signal based on the power with which their signal was transmitted from the base station (BS).

The combination of NOMA with MIMO and mmWave communications was recently reported in [4], [5], where significant improvement in spectral efficiency was demonstrated. Recently, in the context of cellular networks, the integration of NOMA and caching was reported in [7] and [8]. To the best of our knowledge, the only work that investigates the performance of cache-aided NOMA systems in the context of vehicular networks is [9].

Cache-aided NOMA is particularly suitable for the downlink transmission in vehicular networks. This is primarily due to the fact that the SIC operation enables vehicles to alleviate interference from other signals by using their own cache contents. For example, if the file requested by vehicle $V_1$, namely $s_1$, is already cached by another vehicle, e.g., $V_2$, then $V_2$ can remove the interference $s_1$ while decoding its own signal $s_2$. Hence, the performance of a cache-aided NOMA system depends not only on the power allocation of vehicles, but also on the knowledge of the cache contents. Therefore, cache-aided NOMA systems, owing to their ability to remove the interference by exploiting the cached contents, offer a better performance as compared to conventional NOMA systems.

The main contribution of this work is a look-ahead vision that establishes the potential integration of caching and NOMA in the context of vehicular communication networks. In particular, the proposed vision identifies the potential challenges, research opportunities and technological trends that can enable the realization of cache-aided NOMA architecture in vehicular networks. It is further established that the spectral efficiency of the proposed architecture is superior to that of conventional NOMA-based architecture. This is due to the fact that the performance of the former is essentially improved by exploiting the cached contents at vehicles.

## BASIC CONCEPTS OF CACHE-AIDED NOMA

The system description of the proposed cache-aided NOMA system for vehicular networks is depicted in Figure 1. A BS serves, without loss of generality, $N=2$ users (vehicles) in its footprint. Each vehicle is equipped with a cache with finite capacity. The communication between the BS and vehicles is carried out in two phases, namely (a) *content placement phase* (or *cache phase*) and (b) *delivery phase* (or *request phase*). In the content placement phase, each vehicle requests a certain subset of files from the BS, the roadside unit (RSU), or another vehicle, and stores them in its respective cache. Typically, the cache phase is executed during off-peak hours. In the delivery phase, each vehicle requests a new subset of files—possibly during peak hours. The BS serves all requesting vehicles simultaneously, by employing the NOMA technique in downlink. In the next subsections, we explain the steps involved in these two phases in greater detail.

### Caching

As mentioned earlier, in the content placement phase, vehicles cache a subset of popular files from the BS, depending on their popularity profile. Note that the cached subset of files can be different across vehicles and the content replacement can be done with or without partitioning. During the delivery phase, each vehicle requests a new set of files, and simultaneously reports its cached set of files to the BS. The BS serves the newly requested files by transmitting the signals corresponding to these files, using NOMA. In the next section, we explain the two key features of NOMA, that is, the superposition coding at the transmitter and cache-aided successive interference cancellation at the user end.

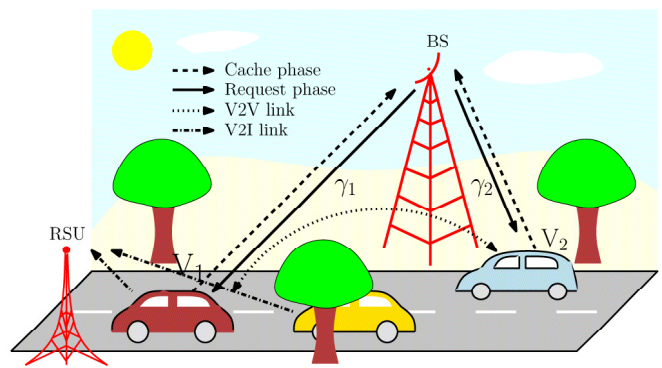

Figure 1. System model of cache-aided NOMA in vehicular networks.

### Superposition Coding

In superposition coding, during the delivery phase, the BS simultaneously transmits different information signals to different vehicles using the same time and frequency resources. The key idea of superposition coding is the power allocation assignment, in which vehicles with poor channel conditions are allocated higher power levels. On the other hand, vehicles with higher channel gains are allocated relatively lower power levels.

As shown in Figure 1, the received signal-to-interference-noise ratios (SINRs) at $V_1$ and $V_2$ are given by $\gamma_1$ and $\gamma_2$, respectively. Since $V_2$ is closer to the BS, i.e., $\gamma_1 > \gamma_2$, the BS assigns a lower power level to $V_1$, i.e., $P_1 < P_2$, where $P_1$ and $P_2$ are the powers allocated to $V_1$ and $V_2$, respectively, and $P_1 + P_2 = P$, with $P$ being the average power available at the BS. Next, the BS transmits the superimposed signal $s = P_1 s_1 + P_2 s_2$, where $s_1$ and $s_2$ correspond to the requests of $V_1$ and $V_2$, respectively. This principle can be extended to more than two vehicles, and the power levels for vehicles are allocated based on their channel gains.

To summarize, the procedures carried out during the content placement phase and superposition coding during the delivery phase, for a general case with $N$ vehicles, are shown in Figure 2. At the receiver end, each vehicle employs a cache-aided SIC decoder which is explained next.

**SIC for Cache-Aided NOMA**

In the conventional NOMA receiver structure, $V_1$ directly decodes its own signal, treating the interference from the signals intended for $V_2$ as noise. On the other hand, the vehicle $V_2$ decodes $s_1$ first, and then subtracts it from its received signal to decode its own signal $s_2$. A similar procedure is carried out with multiple vehicles. In this technique, the vehicles are ordered depending on their received SINR values so that each receiver decodes the strongest signals first, and subtracts them till its desired signal is decoded.

However, the receiver structure following the cache-aided NOMA is slightly different [7], [8]. That is, if the signal intended for $V_1$ is cached at $V_2$ during the content placement phase, then $V_2$ can remove the signal corresponding to $V_1$ directly before decoding its own signal, thereby minimizing the interference. This idea can be extended to the case of multiple vehicles, where a given vehicle can remove the signal intended for another vehicle, provided that it has cached the corresponding file during the content placement phase. Thus, the performance of the cache-aided NOMA system depends on the cache contents of vehicles. This procedure, extended for the general case of $N$ vehicles, is illustrated in Figure 3.

**Generalization of Cache-Aided NOMA**

This subsection discusses how cache-aided NOMA can bridge the two extremes, i.e., NOMA and OMA strategies, by fully treating interference as noise or fully decoding the individual signals. In particular, if a vehicle has not cached any file corresponding to any other vehicle, then the performance of cache-aided NOMA will be similar to that of a conventional NOMA. The other extreme case is where a vehicle has cached all files requested by all other vehicles in its own cache; in this case, the interference at this vehicle can be completely removed. Hence, the performance of cache-aided NOMA on average will be better than that of any single-carrier OMA scheme, e.g., frequency division multiple access. Note that this performance gain is due to the fact that the vehicles are served on the entire bandwidth, as opposed to OMA, where the bandwidth is divided into several orthogonal channels across each vehicle. In the next section, we discuss the performance benefits of the cache-aided NOMA in comparison with the conventional NOMA and OMA schemes.

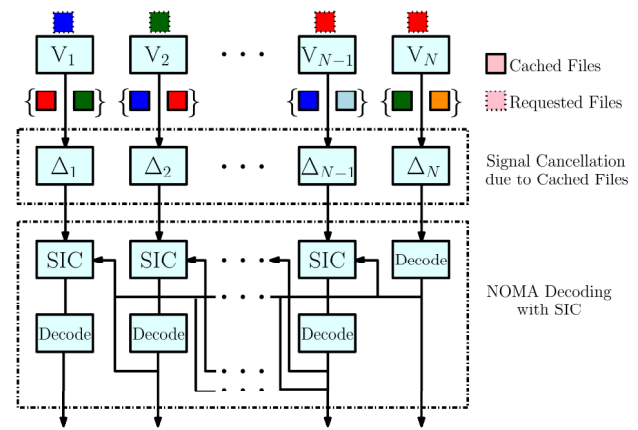

Figure 3. SIC procedure for cache-aided NOMA.

## PERFORMANCE STUDY

In this section, we discuss the performance of the proposed cache-aided NOMA system in terms of probability of successful decoding [8], [9]. We consider a setup with two vehicles, as shown in Figure 1. We assume that a file is successfully decoded if the received SINR is greater than a given threshold, which could be different for different files. For simplicity, we let these thresholds and noise variances to be unity. In a typical vehicular network, the channel gains between vehicles $V_1$ and $V_2$ and the BS can be modelled as independent cascaded Nakagami-*m* distributed random variables, as considered in [10]. Without loss of generality, we let the parameters of the cascaded Nakagami-*m* channels be ((1,1), (2,2)) and ((1,1), (2,2)). The files at the BS are cached at $V_1$ and $V_2$ by following a Zipf distribution with parameter $\zeta$ [3].[3] Therefore, the optimal caching policy is to cache the topmost popular files. Note that as $\zeta$ becomes greater than unity, the popularity profiles of the files at BS have a uniform profile, as opposed to the case when $\zeta$ approaches zero. In the latter case, the popularity of the first file will have an overwhelming probability.

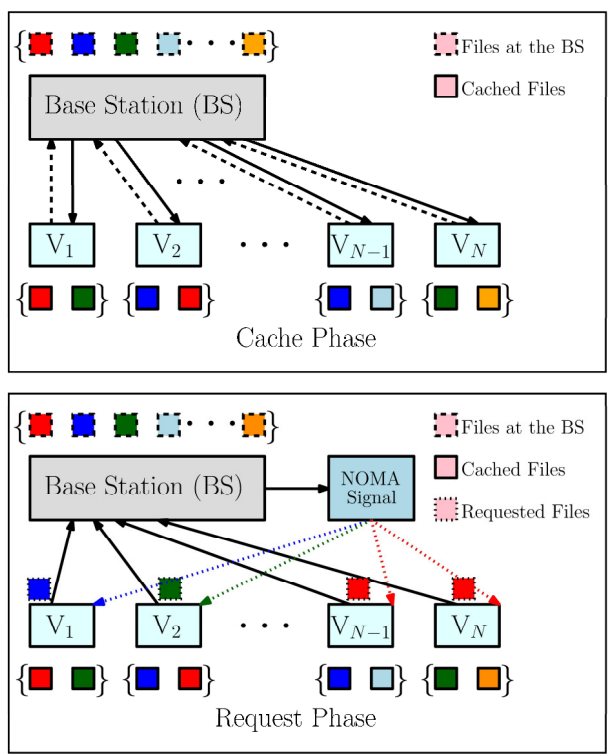

Figure 2. System description during the content placement phase and the superposition coding in the request phase for caching-aided NOMA.

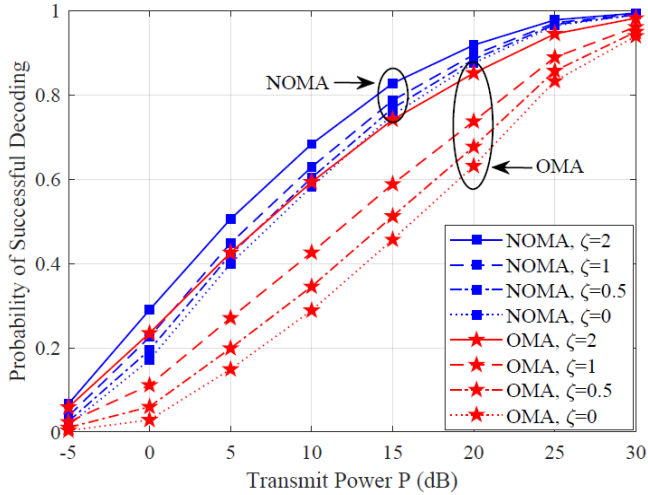

Figure 4. Variation of the probability of successful decoding with SNR for a two-vehicle system for different values of the Zipf parameter, $\zeta$.

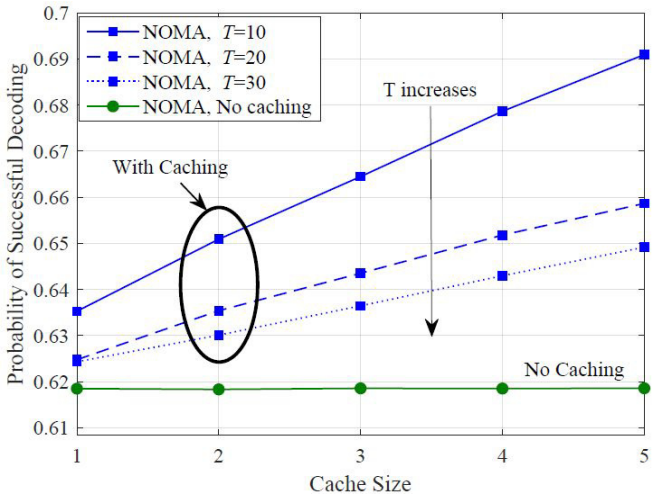

Figure 5. Variation of the probability of successful decoding with cache-size for a two-vehicle system for different values of number of files at the BS, $T$.

Figure 4 compares the performances of cache-aided NOMA and cache-aided OMA strategies for vehicular systems. On the horizontal axis, we plot the total received signal-to-noise ratio (SNR) at vehicles $V_1$ and $V_2$. On the vertical axis, we show the probability of successful decoding, which is the multiplication of probability of the received SINR at $V_1$ and $V_2$ to be greater than a unit threshold. It is seen that as $\zeta$

---

[3] The proposed system can be designed for any meaningful popularity profile.

approaches zero, caching indeed helps in improving the performance of the proposed system.

Figure 5 compares the performance of cache-aided NOMA with conventional NOMA, i.e., without caching. Here, $T$ denotes the total number of files available at the BS. As expected, when the size of the cache at each vehicle increases, the chance that the requested files of $V_1$ and $V_2$ being cached during the content placement phase increases, thereby increasing the performance of cache-aided NOMA. Following the same argument, it can be seen that as $T$ decreases, the performance of cache-aided NOMA increases. Additionally, note that the performance of cache-aided NOMA and conventional NOMA are exactly equal when the cache size is zero, that is, when $V_1$ and $V_2$ are not equipped with storage units.

## FUTURE RESEARCH DIRECTIONS

In this section, we discuss some open research problems and the potential challenges related to the practical realization of the proposed cache-aided NOMA system for vehicular networks.

### Artificial Intelligence and Intelligent Vehicular Networks

Future intelligent vehicles are envisioned to be equipped with hundreds of on-board advanced sensors such as cameras, RADAR, and light detection and ranging, which are expected to generate a massive amount of data. In this regard, machine learning (ML), which is a branch of artificial intelligence (AI), can be used as an effective tool to learn from data, identify unique patterns and finally make proper decisions with minimal human intervention while ensuring reliability and efficiency [11], [12]. ML has recently received significant attention as an emerging field in vehicle automation [13]. Specifically, the state-of-the-art literature has largely focused on: 1) reinforcement learning, which can be used to find efficient routing for communication of data from the BS to individual vehicles; 2) tracking the dynamics of caching in vehicular environments; and 3) the physical-layer aspects of vehicular networks, e.g., optimal power allocation in NOMA.

However, due to high dynamics in vehicular networks and as a result of high mobility, a set of new challenges arises in cache-aided NOMA vehicular networks, which can significantly limit the network performance. Therefore, there is a compelling need to rethink existing wireless design solutions in order to adapt to the fast-changing environments. In the following, we highlight some of the challenges and discuss potential solutions.

*High mobility*: The mobility of vehicles poses several technical challenges which must be adequately addressed for the practical realization of cache-aided NOMA networks. In particular, cache-aided NOMA topology is highly influenced by mobility—affecting resource allocation, scheduling, as well as routing protocol designs. Existing model-based solutions have mainly focused on static and low-mobility scenarios. Therefore, it is imperative to explore new ML-based solutions that can address high-mobility scenarios along with their associated challenges, in order to fully optimize network parameters and offer holistic performance.

*Security in vehicular networks and ML*: Unique aspects such as high mobility, dynamic network topology, and volatility in vehicular networks result in privacy, trust, and security concerns. In particular, the central security problems in a vehicular network are in making the V2X channels secure against cyber-physical attacks. Towards this end, design of robust AI/ML-enabled intrusion detection and anomaly detection algorithms are of paramount importance. In particular, the performance of cache-aided NOMA systems can be notably improved by recognizing these synthesized anomalies with high degree of accuracy through deep learning-driven AI/ML techniques. Within this context, performance studies of cache-aided NOMA systems with physical layer security are of importance.

*Enabling heterogeneous connectivity with stringent quality-of-service (QoS) requirement*: Different types of connections in vehicular networks such as V2V, V2I, and V2X links require diverse and stringent QoS requirements with varying reliability-latency requirements. For example, V2I links require seamless internet access with tolerable limits on the traffic delay, while data on V2V links are delay-sensitive and require high reliability. The conventional wireless communication design techniques are not equipped to meet such diverse QoS requirements, which are further challenged by the high mobility and rapid network dynamics. In this regard, AI/ML techniques, which possess the ability to learn-from and adapt-to the vehicular environment with varying requirements and uncertainties, can help develop satisfactory policies to enhance the performance of cache-aided NOMA systems. Moreover, reinforcement learning can be particularly useful in enabling the interaction with the dynamic environment.

**Vehicular Networks with Large Intelligent Surfaces / Intelligent Reflecting Surfaces**

It is envisioned that the number of connected vehicles, including both ground and aerial vehicles, will increase rapidly in the near future [14]. This would require a paradigm shift in the evolution of wireless technologies. In this respect, THz-band communication is envisioned as a key enabler for the beyond-5G networks, which are expected to combat the explosive growth in data traffic—this can be feasible, given the recent advancements in THz technology. However, the harsh propagation environments still constitute the grand challenge for THz communications.

On the other hand, large intelligent surfaces (LIS) or intelligent reflecting surfaces (IRS) have recently emerged as a novel communication paradigm that is envisioned to revolutionize the way vehicles communicate. LIS is expected to enable seamless connectivity for beyond-5G networks. The LIS/IRS includes a set of passive elements that reflect RF signals to enhance the received SNR and to cover blind spots in wireless communication systems [15]. Rapidly varying channel dynamics and mobility in vehicular networks often degrade the channel quality, which can be enhanced by employing LIS-enabled V2V communication.

The integration of THz technology and LIS is inevitable in the next-generation vehicular networks for several reasons. First, the realization of a complicated phased array structure for beamforming at THz band suffers from high transmission line losses, which on the other hand is significantly low with LIS, since there is no need for transmission lines. This feature, along with the reconfigurable property of LIS, is particularly useful to realize the desired SNR at continuously moving receivers, such as vehicles with dense multipath. Additionally, since the elements of LIS form sharp beams towards specific vehicles, the interference among other vehicles can be mitigated. Hence, LIS-enabled THz communications are envisioned to yield an improved coverage probability, transmission distance and energy efficiency, resulting in an enhanced system performance for cache-aided NOMA in vehicular networks.

Open research problems include, but are not limited to, the optimized deployment of LIS, the development of AI/ML algorithms for reconfiguring LIS and optimizing the network parameters to achieve the optimum performance. In addition, the factors that in general affect the performance of cache-aided NOMA with LIS include: (a) reflection quality, which measures the amount of fractional power reflected by LIS; (b) size, position, and orientation of the LIS, which also contributes to the loss in reflected power; and (c) movement of the surface. The array of reflectors can be installed on a building or on another vehicle, and the physical movement creates weak/strong randomness in the channel gain.

**Visible Light Communications and Free Space Optics in Vehicular Networks**

Visible light communications (VLC) and free space optics (FSO) technologies have received significant research attention in the context of vehicular networks. Typically, optical communication technologies in vehicular networks are suitable for short-range communications—from few tens to hundreds of meters—while they offer a significant performance improvement due to the high SNR, owing to the strong line-of-sight component [13]. Also, the cost of equipment required to realize a VLC/FSO system is relatively low in comparison with cellular systems, since most vehicles are readily equipped with light emitting diodes (LED). Additionally, thanks to the daytime running lights, the VLC/FSO systems can also function during the day. Hence, VLC/FSO has tremendous potential to be a key technology for further enhancing the spectral efficiency in cache-aided NOMA systems for vehicular networks.

Some of the challenges pertaining to the study of cache-aided NOMA in FSO-enabled vehicular networks include the following. First, since VLC/FSO systems are suitable for short-range communications, the channel gain of each user is roughly equal. This strong channel symmetry significantly affects the achievable performance of NOMA systems. Additionally, the VLC systems are sensitive to the non-linearities in optical-to-electrical and electrical-to-optical conversions, which occur due to photodiodes, LED circuitry, and other hardware components such as converters.

**Moving Towards the Sixth Generation (6G)-Enabled Vehicular Networks**

One of the drivers behind the 6G communication systems is the deployment of connected robotics and autonomous vehicles, where the focus is on controlling the latency requirements while enabling transmission of high definition maps. Such controlled rate-reliability-latency trade-off is not available in 5G. The authors in [14] proposed a new communication paradigm termed as *mobile broadband reliable*

*low-latency communications*, which finds its primary applications in autonomous vehicle networks. Additionally, 6G-enabled vehicular networks are expected to evolve towards maximizing volumetric spectral and energy efficiencies (bps/Hz/m$^3$/J), given the three-dimensional physical space spanned by aerial vehicles.

It is predicted that the next-generation of commercial vehicles will be equipped with hundreds of sensors by 2020, which results in data rates in the order of several terabytes per hour. This will pose new challenges in terms of spectral efficiency and massive connectivity for 5G-enabled vehicular networks. In this respect, 6G systems are envisioned to meet these demands by moving from the conventional cellular architecture towards a *cell-less architecture*, through the integration of several communication technologies. The cell-less architecture ensures seamless connectivity without a large overhead due to handovers, while simultaneously maintaining the required QoS. Towards this end, a detailed study on the performance of cache-aided NOMA in a cell-less architecture and the associated challenges is an important research topic.

## CONCLUSION

This article has presented a cache-aided NOMA architecture for vehicular networks. As a main highlight, it is discussed that the performance of the proposed system not only depends on the channel conditions of each vehicle, but also on the cached contents at each vehicle. A critical study on the design and implementation aspects of the proposed system, along with a detailed state-of-the-art review has enabled us to realize the potential of the integration of the two technologies for vehicular networks, namely caching and NOMA. While there are associated advantages and disadvantages, the integration of these two techniques, along with other promising technologies discussed in the research directions, are expected to pave the way for the practical realization of the near-future vehicular wireless communication systems and beyond.

## ACKNOWLEDGMENT

This work was supported by the Information and Communication Technology Fund. The work of O. A. Dobre has been supported in part by the Natural Sciences and Engineering Research Council of Canada (NSERC).

**SANJEEV GURUGOPINATH** received the Ph. D. degree in electrical communication engineering from the Indian Institute of Science, Bangalore, India. He is currently a Professor in the Department of Electronics and Communication Engineering at PES University, India. His research interests are in the broad areas of signal processing and wireless communications. He is the recipient of Best Paper Awards at IEEE INDICON 2016 and ICEECCOT 2019.

**SAMI MUHAIDAT** [M'07, SM'11] received the Ph.D. degree in electrical and computer engineering from the University of Waterloo, Ontario, in 2006. From 2008-2012 he was an Assistant Professor in the School of Engineering Science, Simon Fraser University, Canada. He is currently a Full Professor at Khalifa University. His research focuses on physical layer aspects of wireless communications. He is currently an Area Editor for IEEE Transactions on Communications. Previously, he served as a Senior Editor for IEEE Communications Letters.

**YOUSOF AL-HAMMADI** received the Ph. D. degree in Computer Science and Information Technology from the University of Nottingham, U. K., in 2009. He is currently a Director of Graduate Studies, Khalifa, UAE. His main research interests are in the area of information security, which includes intrusion detection, botnet/bots detection, viruses/worms detection, artificial immune systems, machine learning, RFID security, and mobile security.

**PASCHALIS C. SOFOTASIOS** [M'12, SM'16] received the M.Eng. degree from Newcastle University, U.K., in 2004, the M.Sc. degree from the University of Surrey, U.K., in 2006, and the Ph.D. degree from the University of Leeds, U.K., in 2011. He has held academic positions at the University of Leeds, U.K., and the Khalifa University, UAE, where he currently serves as an assistant professor. His research interests include broad areas of digital and optical wireless communications. He served as an Editor for IEEE Communications Letters.

**OCTAVIA A. DOBRE** [M'05, SM'07, F'20] is a Professor and Research Chair at Memorial University (Canada), which she joined in 2005. Octavia received the Dipl. Ing. and Ph.D. degrees from the Polytechnic University of Bucharest (formerly Polytechnic Institute of Bucharest), Romania in 1991 and 2000, respectively. She was a Royal Society Scholar and a Fulbright Scholar. She is a fellow of the Engineering Institute of Canada and a Distinguished Lecturer of the IEEE Communications Society. She is the recipient of diverse awards, including Best Paper Awards at IEEE ICC, IEEE Globecom and IEEE WCNC. Her main research areas are NOMA, full-duplex, signal identification, as well as optical and underwater communications.